\documentclass[submission,copyright,creativecommons]{eptcs}
\providecommand{\event}{MBT 2012}
\usepackage{amsmath}
\usepackage{amssymb}
\usepackage{amsthm}
\usepackage{graphicx}
\usepackage{url}
\usepackage{xspace}

\newcommand{\focus} {AutoFocus \xspace}
\newcommand{\N} {\mathbb{N}}
\newcommand{\Hist} {\mathbb{H}}
\newtheorem{definition}{Definition}

\title{Reusing Test-Cases on Different Levels of Abstraction in a Model Based Development Tool}
\author{
Jan Olaf Blech \quad\qquad Dongyue Mou \quad\qquad Daniel Ratiu
\email{\quad \{blech,mou,ratiu\}@fortiss.org}
\institute{fortiss GmbH}
}
\def\titlerunning{Reusing Test Cases}
\def\authorrunning{Blech, Mou, Ratiu}

\begin{document}
\maketitle
\begin{abstract}
Seamless model based development aims to use models during all phases of the development process of a system. 
During the development process in a component-based approach, components of a system are described at qualitatively differing abstraction levels: during requirements engineering component models are rather abstract high-level and underspecified, while during implementation the component models are rather concrete and fully specified in order to enable code generation. 
An important issue that arises is assuring that the concrete models correspond to abstract models. In this paper, we propose a method 
to assure that concrete models for system components refine more abstract models for the same components. In particular we advocate a framework for reusing test-cases at different abstraction levels. Our approach, even if it cannot completely prove the refinement, can be used to ensure confidence in the development process. In particular we are targeting the refinement of requirements which are represented as very abstract models. Besides a formal model of our approach, we discuss our experiences with the development of an Adaptive Cruise Control (ACC) system in a model driven development process. This uses extensions which we implemented for our model-based development tool and which are briefly presented in this paper. 
\end{abstract}

\section{Introduction}
Formally defining relations between different models used in the development process of an entire system or a component is an important prerequisite to apply techniques that give confidence in the refinement process. Lifting properties -- in our case single test-cases -- to work on more refined models is one technique towards gaining this confidence. It is in general much easier and scalable to perform than, e.g., using formal verification techniques, and thus more convenient to use for a wider range of applications and developers that are not trained in the usage of formal methods. 

Abstract models during an early phase in the software development process can be used to capture single requirements directly.
 Such models of single requirements concentrate on distinct aspects of a system and thereby are rather small. These models may represent functional requirements and are relatively easy to validate using test-cases and a simulation environment.
At a later stage in the development process we can  have a model that is more refined and may describe the entire system functionality with the interplay of several fine-grained requirements. A challenge is to use the existing test-cases to get confidence that the concrete implementation models refine the abstract requirement models.

In this work, we regard the relation between abstract models that may correspond to single requirements, and  more concrete models that can already contain enough information to automatically generate an implementation. 
We regard the transformation of existing test-cases for the abstract models into test-cases for the concrete ones.

The initial motivation for this work -- that we faced in a project -- is a situation  where manufacturers of larger systems (e.g., a car) give an abstract specification with abstract test-cases to sub-contractors that are responsible for developing a distinct component (e.g., a control system used in the car). Here, at least the provided test-cases for the abstract specification must be respected by the implementation that the sub-contractor provides.

We present a formal framework for relating test-cases, report on our implementation experiences and regard a case-study on Adaptive Cruise Control (ACC) systems.
Requirements ACC on  systems are standardized in the ISO 15622 standard \cite{iso15622}. In this paper, we present some requirements and a model that captures the implementation and evaluate our test-case transformation using this case-study. The featured implementation and case study is integrated into our AutoFocus 3
\cite{autofocus} model-based development environment and its modeling language.

Our paper features the following contributions:
\begin{itemize}
\item A formal framework for relating models of components for different abstraction levels which aims especially at concretizing test-cases for AutoFocus 3.
\item An implementation of the framework in AutoFocus 3.
\item We exemplify our approach using a simplified variant of an ACC system.
\end{itemize}
Thus, we are addressing the issue of transforming existing test-cases and relating abstract and concrete models with each other. Derivation of new test-cases and evaluation of test-case quality is not subject to this paper.

\subsection{Related Work}
Relations between properties on abstract and concrete system representations have been studied comprehensively for temporal logics formula, e.g., in \cite{LGSBB95}. Our test-cases (test + expected result) may be regarded as properties if classified in the terms of this paper.

Abstractions in the context of model-based testing has been studied in \cite{pretschner04}. Here, in contrast to our work, different abstraction levels of models are studied regarding suitability for deriving test-cases for a final implementation.

An early  formal framework for relating test-cases, interactive systems and abstractions is described in \cite{aichernig01}. Here, test-cases are regarded as contracts, which is similar to our view. 

The usage of program analysis techniques on abstract system representation has been studied in the context of abstract testing \cite{CousotCousot-SSGRR-00}, a kind of method for handling testing in an abstract interpretation framework.

The work presented in \cite{yorsh06} presents a combination of abstract interpretation and model checking for testing. Here, somehow opposite to our work, concrete test-cases are abstracted for abstract system representations.

In addition to test-case abstraction and concretization, a large body of research has been done on other aspects of model-based testing. For a comprehensive overview and classification  we refer to \cite{MBTtaxonomy}.

\subsection{Overview}
In Section~\ref{sec:focus} we describe our model-based development tool AutoFocus 3 and the semantics of the used language. 
The framework for relating models and transforming test-cases is presented in Section~\ref{sec:framework}. 
Our case study and a small evaluation is presented in Section~\ref{sec:acc}.
Section~\ref{sec:conc} features a conclusion.

\section{Our Model Based Development Environment}
\label{sec:focus}

In this section we present the model based development tool: AutoFocus 3 that we use in the context of this work. In particular we present the semantics of the modeling language. 
\subsection{AutoFocus Semantics}
\label{sec:a3sem}
Here we present a formal definition of our modeling language AutoFocus a dialect of FOCUS \cite{2001_broy_specification_and_development_of_interactive_systems}. We follow the description given in \cite{focusTR}

In AutoFocus a system and its components are described by a \emph{stream processing function}, which defines its syntactic interface as well as its behavior.
Furthermore, the \focus approach offers composition operators which allow to derive a larger system (the composed system) out of modularly defined functions.

\subsubsection{Streams}
Basically, the \focus theory is based on the idea of timed data streams which are used to model the asynchronous interaction between a function and its environment.
Streams represent histories of communication of data messages in a given time frame.
Intuitively, a timed (data) stream can be thought of as a chronologically ordered sequence of data messages.

\begin{definition} [Timed Stream]
Given a set $M$ of data messages, we denote a timed stream of elements from $M$ by a mapping
\[
s:\N\rightarrow M\text{.}
\]

For each time $t \in \N$, $s(t)=s.t$ denotes the message communicated at time $t$ in a stream $s$ and 
$s\downarrow t$ the prefix of the first $t$ messages in the timed stream $s$, i.e., the messages communicated until (and including) time $t$.\footnote{The theory is originally defined for stream processing functions $s:\N_{+}\rightarrow M^\ast$, which assign a sequence of messages to each time interval. In order to keep the paper at hand as understandable as possible, we decided for the simplification that only one message can be communicated within each time interval. A proper description of the original theory can be found in~\cite{2001_broy_specification_and_development_of_interactive_systems,2007_broy_formal_model_of_services}}
\end{definition}

We have chosen so-called timed data streams that allow us to flexibly include the timing issues of functions whenever required. 
We base our approach on a simple notion of discrete time: we assume a model of time consisting of an infinite sequence of time intervals of equal length. 
Thus, time can be simply represented by the natural numbers $\N_{+}$.
In each time interval a message $m \in M$ can be transmitted.

An exemplary timed stream $s$ over the data set $\mathbb{B}=\{0,1\}$ is defined by the function $\forall t\in\N:s(t)=1$. 
This means that in each time interval the stream contains the value $1$, i.e., $s = (1~1~1~\ldots).$

\subsubsection{Input/Output Channels and Channel Histories}
Every stream processing function is connected to its environment by channels. 
The channels of a stream processing function are divided into disjoint sets of input channels $I=\{i_1, \ldots i_n\}$ and output channels $O=\{o_1, \ldots o_n\}$.
Channels are used as identifiers for streams.
\begin{figure}[htbp]
	\centering
		\includegraphics[width=0.50\textwidth]{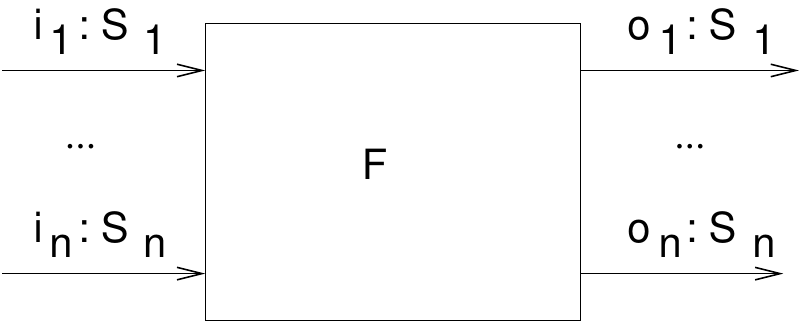}
	\caption{Graphical representation of a FOCUS function and its typed I/O channels}
	\label{fig:focus_function}
\end{figure}

With every channel $c$, we associate a data type $\mathit{Type}(c)$ indicating the set of messages sent along this channel. 
To that end, we define the channel type by the following function:
\[
\mathit{Type} : C \to \mathtt{Type},
\]
which maps each channel $c \in C$ to a data type $t \in \mathtt{Type}$ from the set of all possible data types $\mathtt{Type}$.

To describe the function's communication with its environment, each channel is associated with a stream which represents the messages communicated over this channel (cf.\ Figure~\ref{fig:focus_function}).
A mapping that associates a stream to any channel from a set of channel $C$ is called \emph{(channel) history} of $C$.
\begin{definition}[Channel History]
Let $C$ be a set of typed channels. A channel history $h$ is a mapping
\[
h : C \to (\N \to M),
\]
such that $h(c)$ is a stream of the type $\mathit{Type}(c)$ for each $c \in C$.

The set of all histories over the set of channels $C$ is denoted by $\Hist(C)$.
\end{definition}

\subsubsection{Specification of Stream Processing Functions}
The black-box specification of a stream processing function consists of a syntactic interface and its semantics.

A stream processing function is connected to its environment exclusively by its syntactic interface consisting of input/output channels.
This interface indicates which types of messages can be exchanged.
\begin{definition} [Syntactic Interface]
The syntactic interface of a function is denoted by
\[
(I \blacktriangleright O),
\]
where $I$ and $O$ denote the sets of typed input and output channels, respectively.
\end{definition}

For a function with syntactic interface $(I \blacktriangleright O)$, the set of all syntactically correct history pairs is denoted by 
\[
\Hist(I) \times \Hist(O)\text{.}
\]
However, the syntactic interface tells nothing about the interface behavior of the function.

The behavior (semantics) of the stream processing function is given by the mapping of histories of the input channels to histories of the output channels.
Thereby, we distinguish between total and partial functions.
While the behavior of a total function is defined for \emph{all} syntactically correct inputs, the behavior of a partial functions is defined for a \emph{subset} of the inputs.

\begin{definition}[Semantics Relation]
\label{def:semantic_function_total}
The semantics of a total stream processing function with syntactic interface $(I \blacktriangleright O)$
is given by a relation
\[
F:~\Hist(I)\rightarrow\mathcal{P}(\Hist(O))
\]
that fulfills the following timing property for all its input histories.

Let be $x_1, x_2 \in \Hist(I), y_1, y_2 \in \Hist(O)$, and $t \in \N_{+}$. The timing property is specified as follows:
\[
x_1\downarrow t=x_2\downarrow t\Rightarrow \{y_1\downarrow(t+1):y_1\in F(x_1)\}=\{y_2\downarrow (t+1):y_2\in F(x_2)\}\text{.}
\] 
\end{definition}

By mapping into the power-set of $\Hist(O)$, Definition~\ref{def:semantic_function_total}  allows to specify \emph{nondeterministic} behavior.
For an input history, there is a set of output histories that represent all possible reactions of the function to the input history. 
If a function defines exactly one output history for every input history, the function is called deterministic; if the set of output histories has several members for some input history, then the function is called nondeterministic.

The timing property expresses that the set of possible output histories for $F$ for the first $t+1$ intervals only depends on the inputs of the first $t$ time intervals.
In other words, the processing of messages within a function takes at least one time interval.
Functions that fulfill this timing property are called time-guarded or \emph{strictly causal}.
Strict causality is a crucial prerequisite for the composability of functions.

If we replace the expression $(t+1)$ by $t$ in Definition~\ref{def:semantic_function_total} above (i.e., the outputs in the first $t$ intervals depend on the inputs in the first $t$ intervals), messages are processed by the function without time delay.
Such functions are called \emph{weakly causal}.

Stream processing functions are used to represent components in our terminology. They can contain other stream processing functions thereby forming composed components. 

Moreover, it is important to notice that stream processing functions in the AutoFocus realization of FOCUS are defined as finite automata comprising internal states and not in a functional way working on entire streams. The definition is rather element wise, thus, consuming one element for each input channel at a time and generating an output element for each channel. 

For further information, we refer to \cite{2001_broy_specification_and_development_of_interactive_systems,2007_broy_formal_model_of_services} for a discussion of partial stream processing functions, weak and strong causality, and a deeper discussion of the semantics of FOCUS.

\subsection{Tool Integration and Proposed Usages}
The AutoFocus modeling language has been implemented in a model based development tool: AutoFocus 3 \cite{autofocus}.  AutoFocus 3 supports structuring textual requirements and expressing them through models. It allows the graphical modeling of system components, their functionality and deployment aspects.  It comprises code generation and integrates formal verification and validation tools. AutoFocus 3 is based on Eclipse RCP framework and comes with a plug-in mechanism for new functionality. 
 
AutoFocus 3 aims at the development of embedded systems, e.g., in the automotive industry. It can be used for the entire development process.
In a typical workflow with AutoFocus 3, project partners specify requirements in a textual way using AutoFocus, create corresponding models and establish test-cases based on these requirements. In the course of the project, partners provide more detailed models and an implementation. It is important, that the test-cases for the models representing higher abstraction levels of a system component can be transformed into test-cases for the more refined models. 
\section{An Abstraction Framework for Models and Test-cases}
\label{sec:framework}
In this work, we want to relate abstract and concrete models and lift test-cases from more abstract models to more concrete ones. Characteristic for our framework are the facts that both abstract and concrete models are given in the same modeling language. In AutoFocus, a model does typically represent a (potentially composed) component. In AutoFocus a component's only means of interaction with its environment is  via its channels. Thus,  we do not have to deal with the possibility of additional side effects which facilitates a formal relation.

An important issue is the correctness of the lifting and if the refinement between models has been done correctly. Here we present basic definitions for test-cases and two formalisms for relating abstract and concrete models which we have applied and evaluated in our model-based development environment:
\begin{enumerate}
\item A formalism for relating inputs of test-cases of their expected results using mathematical relations.
\item A formalism for relating the stream domains of abstract and concrete components using Galois connections.
\end{enumerate}

\subsection{Formal Definition of Test-cases}
Here we present some basic definitions for test-cases. Test cases comprise test-input and expected results.

\begin{definition}[Test-input]
A test-input for an AutoFocus Component with   syntactic interface $(I \blacktriangleright O)$ is a function $c \mapsto s$ with $c \in I$ and $s$ a correctly typed stream (cf. Section~\ref{sec:a3sem}). 
\end{definition}

\begin{definition}[Test expected-results]
An expected-result of a test-case for an AutoFocus component with   syntactic interface $(I \blacktriangleright O)$  is a set of tuples $(c,s)$ with $c \in O$ and $s$ a correctly typed stream. 
\end{definition}
For the deterministic systems that we regard  in this paper expected results can also be written as functions from output channels to streams.

\begin{definition}[Test-case]
A test-case for an AutoFocus component 
is a tuple of test-input and expected-result .
\end{definition}

\subsection{Relating Test-inputs and Expected-results}
\label{sec:correctnessRIRO}
\begin{figure} 
\begin{center}
	\includegraphics[width=0.65\textwidth]{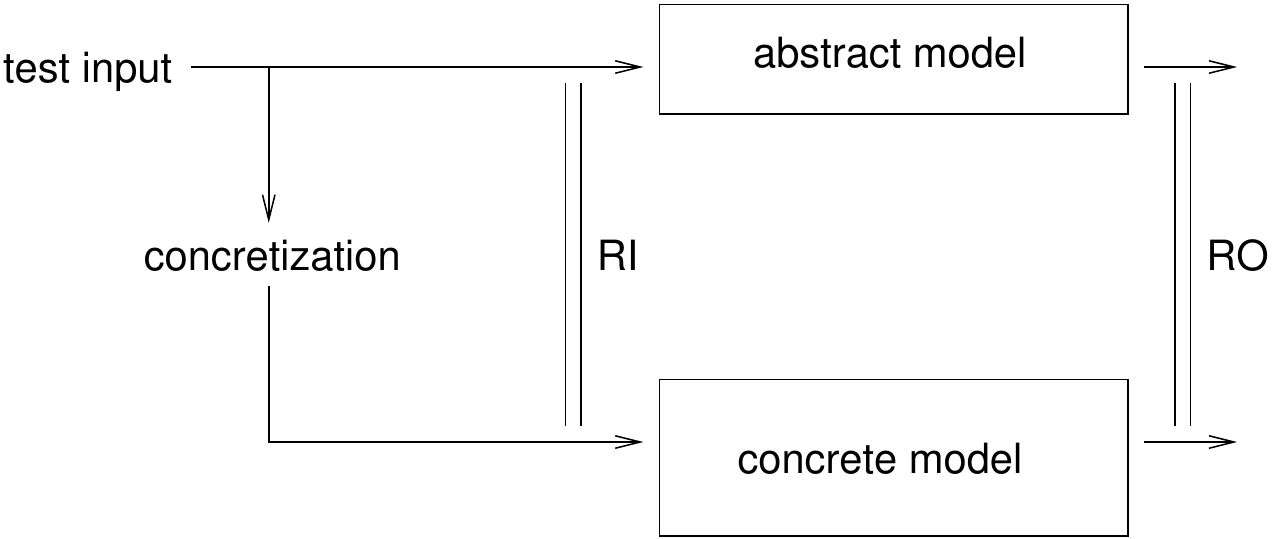}
\end{center}
\caption{Our testing scenario}
\label{fig:abso}
\end{figure}

Figure~\ref{fig:abso} shows our first formalism for relating models and test-cases: 
As a prerequisite, we establish two abstraction relations {\sf RI} and {\sf RO} which capture the relation between abstract and concrete component and compare the inputs of the two components and their results with each other.  These can be stated in a formal way. Then we create an AutoFocus component which transforms abstract test-cases into concrete ones.

When such an existing test input for an abstract model is transformed into a concrete one we compare these test inputs and the expected-results using the relations thereby validating the correctness of test-case transformation. 

Formally, we regard an abstract component with semantics function $C_a$ and syntactic interface $(I_a \blacktriangleright O_a) $ featuring typed input channels $I_a$ and output channels $O_a$. The concrete model has a semantics functions $C_c$ and syntactic interface $(I_c \blacktriangleright O_c)$ featuring input channels $I_c$, and output channels $O_c$. 
The relation {\sf RI} has the type $I_a \times I_c \rightarrow bool$, {\sf RO} : $O_a \times O_c \rightarrow bool$. 
These two relations have to be created for every pair of abstract and concrete models. 

The formal definition of correctness of a test-case transformation is stated in the following definition.
\begin{definition}[Corresponding test-inputs]
Two test-inputs $t_a$, $t_c$ are corresponding with respect to input and output correctness relations {\sf RI}, {\sf  RO} for abstract and concrete \focus components with semantic functions $C_a$, $C_c$ typed accordingly iff
\begin{center}
$\mathsf{RI} (t_a,t_c) \rightarrow \mathsf{RO} (C_a(t_a),C_c(t_c))$ 
\end{center}
\end{definition}

Here, we assume that it is relatively easy to ensure that these the relations have been stated correctly. Checking that the relations have been done correctly can be done either manually or by implementing an AutoFocus component which performs the check and emits a boolean stream of check values.
 
\subsection{Relating the Domains of Components Using Galois Connections}
\label{sec:galois}
A second way to relate abstract and concrete components with each other is to formalize a relationship between their domains. 

Both abstract component (syntactic interface $(I_a \blacktriangleright O_a)$) and concrete component (syntactic interface $(I_c \blacktriangleright O_c)$) operate on domains of streams. When regarding the abstract input and output domains together as a single abstract domain and the concrete input and output domains as a single concrete domain and in case the abstract component is refined by the concrete one it is reasonable to relate abstract and concrete domain with each other  via a Galois connection. Galois connection are used to represent refinements between different domains and  capture their relation in a formal way (cf., e.g., \cite{CousotCousot-SSGRR-00}).

This implies the existence of two functions $f : I_c \cup O_c \rightarrow I_a \cup O_a$ and $g : I_a \cup O_a \rightarrow I_c \cup O_c$. Intuitively $f$ lifts sets of concrete input and output streams to abstract ones and $g$ does it the other way round. Furthermore,  in order to illustrate the refinement between the two domains the functions have to fulfill the following conditions:
\begin{center}
$\forall \ T^{io}_c \ T^{io}_a \ . \ f(T^{io}_c ) \subseteq T^{io}_a$ iff $ T^{io}_a \subseteq g(T^{io}_c  ) $ 
\end{center}
The functions $f$ and $g$ capture the nature of the refinement. 
We formalize them in order to gain a precise description of the concretization  allowing a transformation of test-cases.
 
In our proposed scenario we regard the transformation of abstract test-cases to concrete ones. The process of concretizing a test-case in general allows for multiple concrete test-cases for a given abstract one. Once we have established the function $f$, we formalize a parameterized family (parameter $p$) of functions $f^{-1}_p$ to perform this task for the test-input data. 
At least the part of $f^{-1}_p$ is implemented as an AutoFocus component that covers the test-input. Its semantics function is denoted $F^{-1}_p$. In case of a restriction to the input streams all instantiations of the component have the syntactic interface  $(I_a \blacktriangleright I_c)$.

In accordance with the Galois connection $f$ is in general too restrictive for comparing the expected-results of test-cases, thus, we use $g$ or $g^{-1}$ for this purpose.

\subsection{Implementing Test-case Transformation and Checking}
For a practical implementation in our tool the concretization is implemented using an AutoFocus component which corresponds to the relation {\sf RI}. Checking the results of test-inputs is done using the relations {\sf RO}. Both {\sf RI}  and {\sf RO} and the related test-case concretization can be constructed based on the functions $f$ and $g$ and their inverses.

\begin{figure} 
\begin{center}
	\includegraphics[width=0.75\textwidth]{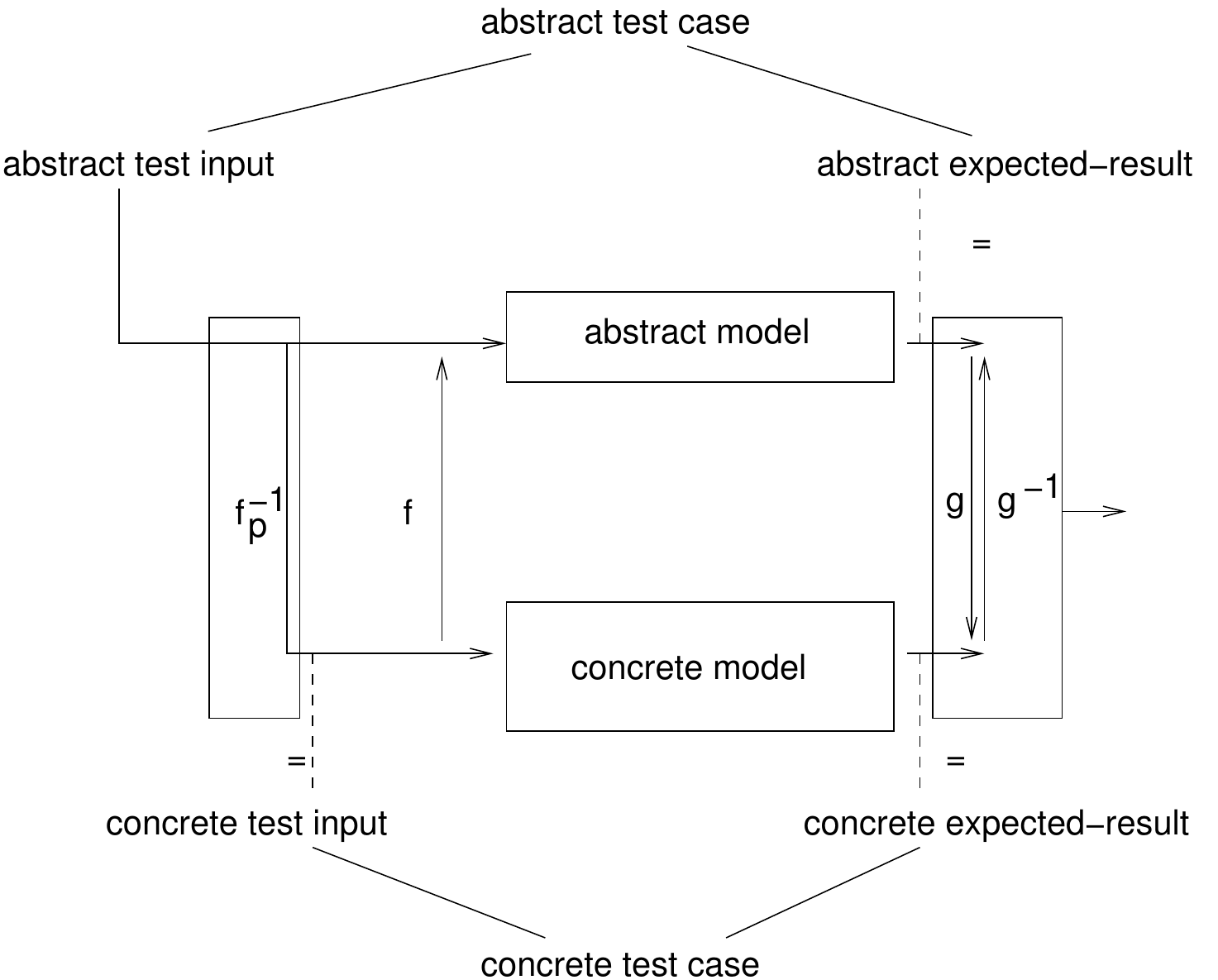}
\end{center}
\caption{Realization of the testing scenario}
\label{fig:absofg}
\end{figure}

More concretely, for checking the output relation {\sf RO} we realize an AutoFocus component and take a semantic function based on $g$ or $g^{-1}$ (if $g^{-1}$ exists). The realization is shown in Figure~\ref{fig:absofg}.
For convenience reasons, we embed it into a component which just checks that the output streams are in relation. Thus, this component has the syntactic interface $(O_a \times O_c \blacktriangleright {\mathit{bool stream}}) $ that checks whether the output streams are compatible and emits a boolean value accordingly. The bool stream must be interpreted in a way, such that a single false occurring in the streams turns the entire comparison into false.

\subsection{An Example}
\begin{figure} 
\begin{center}
	\includegraphics[width=0.4\textwidth]{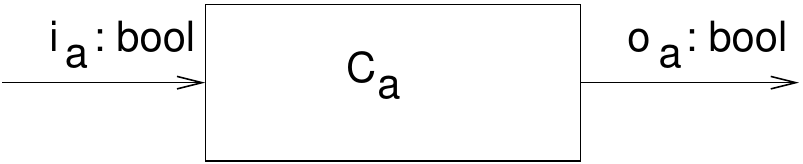}
\end{center}
\caption{An abstract component}
\label{fig:abscomp}
\end{figure}

\begin{figure} 
\begin{center}
	\includegraphics[width=0.4\textwidth]{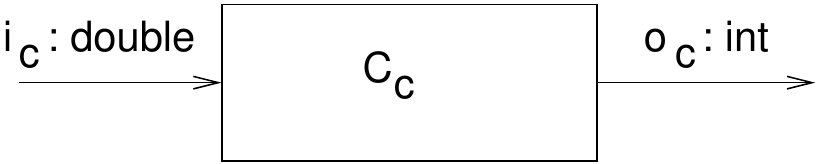}
\end{center}
\caption{A concrete component}
\label{fig:conccomp}
\end{figure}

Consider the abstract component in Figure~\ref{fig:abscomp} and a concretization in Figure~\ref{fig:conccomp}. The concrete component shall encode and abstract a 64 bit floating point value -- e.g., an input from a sensor -- into an 8 bit integer representation and the fact whether it is a positive or a negative number shall be preserved. 
One basic requirement specified by the abstract component could be that positive input values (including $0$) are encoded into positive output values (including $0$) and negative input values are encoded as negative output values.  
\paragraph{Establishing the relation based approach}
For using the first approach, we can establish the two relations defined on sets of streams for the channels $i_c$, $i_a$, $o_c$, $o_a$: 
\begin{center}
{\sf RI}  $(I_a,I_c) = \forall \ t \ \in \ \N_+ \ , \ ( \ I_a(i_a).t \  \wedge \ I_c(i_c).t \   \geq  \ 0 \ ) \ \vee \  ( \ \neg I_a(i_a).t  \ \wedge  \ I_c(i_c).t \ \le \ 0 \ ) $ 
\end{center}
\begin{center}
and 
\end{center}
\begin{center}
{\sf RO} $(O_a,O_c) = \forall \ t \ \in \ \N_+ \ , \ ( \ O_a(o_a).t \  \wedge \ O_c(o_c).t \   \geq  \ 0 \ ) \ \vee \  ( \ \neg O_a(o_a).t  \ \wedge  \ O_c(o_c).t \ \le \ 0 \ ) $ 
\end{center}
A test case with $[true;false;true]$ as input stream for the {\sf i} channel will deliver the output $[true;false;true]$ for the abstract component.
A correct concretization may be $[2.5;-3.6;0.3]$ and might generate the (correct) output $[2;-4;0]$ on the concrete component.

\paragraph{Using the Galois connection approach}

For the second approach we relate the two models by using the functions $f$ and $g$.

For example $f$  can be defined as:
\begin{center}
$f (T^{io}_c ) = \{ (\mathit{abs}(i_c),\mathit{abs}(o_c)) | (i_c,o_c) \in T^{io}_c \}$
\end{center}
$\mathit{abs}$ is a function that works on streams and abstracts the floating point and integer values to bools, in a way corresponding to the example  above. While  
\begin{center}
$f(\{([2.5;-3.6;0.3],[2;-4;0])\}) = \{([true;false;true],[true;false;true])\}$ 
\end{center}
the opposite way allows in general multiple concretizations, 
\begin{center}
$f^{-1}_p (T^{io}_a ) = \{  p(i_c,o_c) |  (i_c,o_c) \in T^{io}_c \} $
\end{center}
e.g. for a concrete implementation $f$,
\begin{center}
$f^{-1} (  \{([true;false;true],[true;false;true])\} ) = \{([2.5;-3.6;0.3],[2;-4;0])\}$
\end{center}

In this case $g$ can be realized in the following way
\begin{center}
$g (T^{io}_a ) = \{ (i_a,o_a) | \exists \ i_c \ o_c\ (\mathit{abs}(i_c),\mathit{abs}(o_c)) \in T^{io}_a \}$
\end{center}
e.g., 
\begin{center}
$g (  \{([true;false;true],[true;false;true])\} ) =  \{ ([2.5;-3.6;0.3],[2;-4;0]), ([2.4;-3.8;0.4],[2;-4;0]), ([2.5;-3.6;0.3],[2.5;-4.4;0]), ... \} $
\end{center}

Note that in general $f^-1 \subseteq g$ holds. 
The transformation of test-inputs and checking the correspondence of the results can be done using AutoFocus components build from $f^{-1}_p$ and $g$ functions. 

\paragraph{Complexity of test-case comparison and limitations} 
The proposed method uses entire abstract and concretized streams. Thus, components, that realize the abstraction and concretization functions may contain internal states if they are defined in an element-wise way on the streams. 

Another characteristic is that input and output streams are regarded independently. This is justified by the fact that the relation between input and output streams is covered by the abstract model and correctness of test-cases and the concrete model is always regarded with respect to this abstract model.

\section{ACC Case Study}
\label{sec:acc}

We have chosen a simplified variant of an Adaptive Cruise Control System(ACC) as the case study. The ACC system controls the speed of the car and takes a desired default speed, a distance to the car in front and inputs like pressing the gas or brake pedal into account. The entire development process is carried out using AutoFocus 3. In particular, textual requirements are structured and specified, models are build from these requirements and test-cases are generated for these models.  Several refined models for components are created in the development process using  AutoFocus 3 until finally an implementation is created. Here, we start with requirements of an ACC systems which are taken from the ISO 15622 standard \cite{iso15622} ``Intelligent transport systems -- Adaptive Cruise Control systems -- Performance requirements and test procedures''.

\begin{figure} 
\begin{center}
	\includegraphics[width=0.95\textwidth]{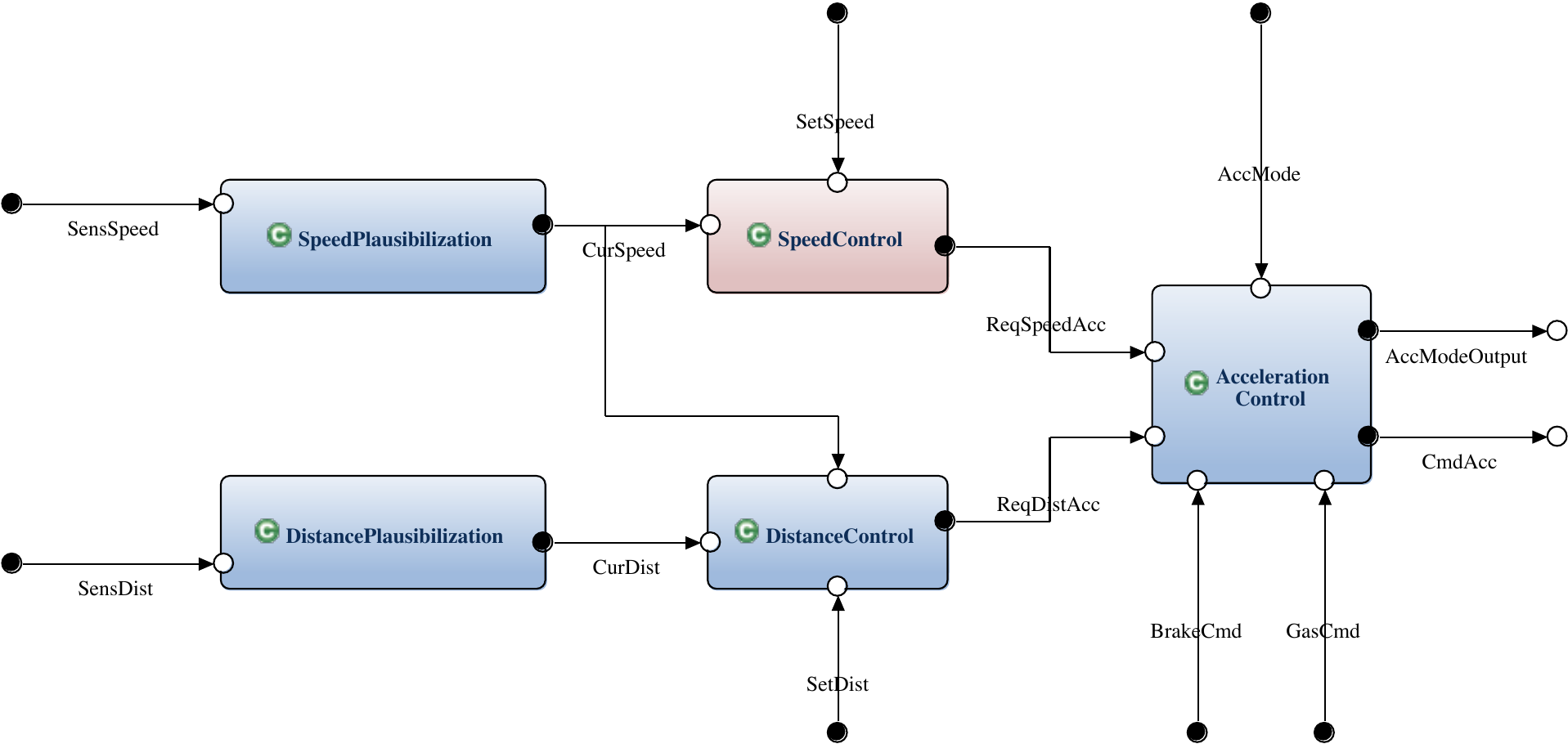}
\end{center}
\caption{An ACC model in \focus}
\label{fig:acc-afm}
\end{figure}

Figure~\ref{fig:acc-afm} shows the top-level component structure including input and output streams of the ACC that is regarded in this paper as realized in AutoFocus.
This model works on several input streams: {\it SensSpeed} representing current speed values, {\it SensDist} representing the distance to the next car in front of the car with the ACC. Furthermore, streams that represent user input to set an acceleration mode ({\it AccMode}), a desired speed value ({\it SetSpeed}), a desired distance to the next car ({\it SetDist}), and pressing the brake or gas pedal ({\it BrakeCmd} and {\it GasCmd}) are included. Output streams comprise values for showing the current Mode ({\it AccModeOutput}) and an acceleration command which can be negative when braking ({\it CmdAcc}) or positive when an acceleration is considered favorable by the ACC.

The ACC model consists of 5 components. The SpeedPlausibilisation component (left top) and DistancePlausibilisation component  (left bottom) measure takes the input speed and distance and delivers the calibrated values to the speed control component (middle top) and distance control component (middle bottom). Both components compute the suitable acceleration values according to the given speed and distance and send the results to the acceleration control component (right). When the acceleration control component is turned on and thus in active mode, it decides whose value should be respected. When it is switched to stand-by mode, both values will be discarded and only user's inputs are being processed. For safety consideration, drivers can provide their inputs any time (accelerate or brake) and this forces the acceleration control component to  switch itself to stand-by mode.

Remember that components are stream processing functions and are composed from subcomponents. A simplified version of the ACC system can be downloaded with the standard AutoFocus 3 distribution \cite{autofocus}.

\begin{figure} 
\begin{center}
	\includegraphics[width=0.8\textwidth]{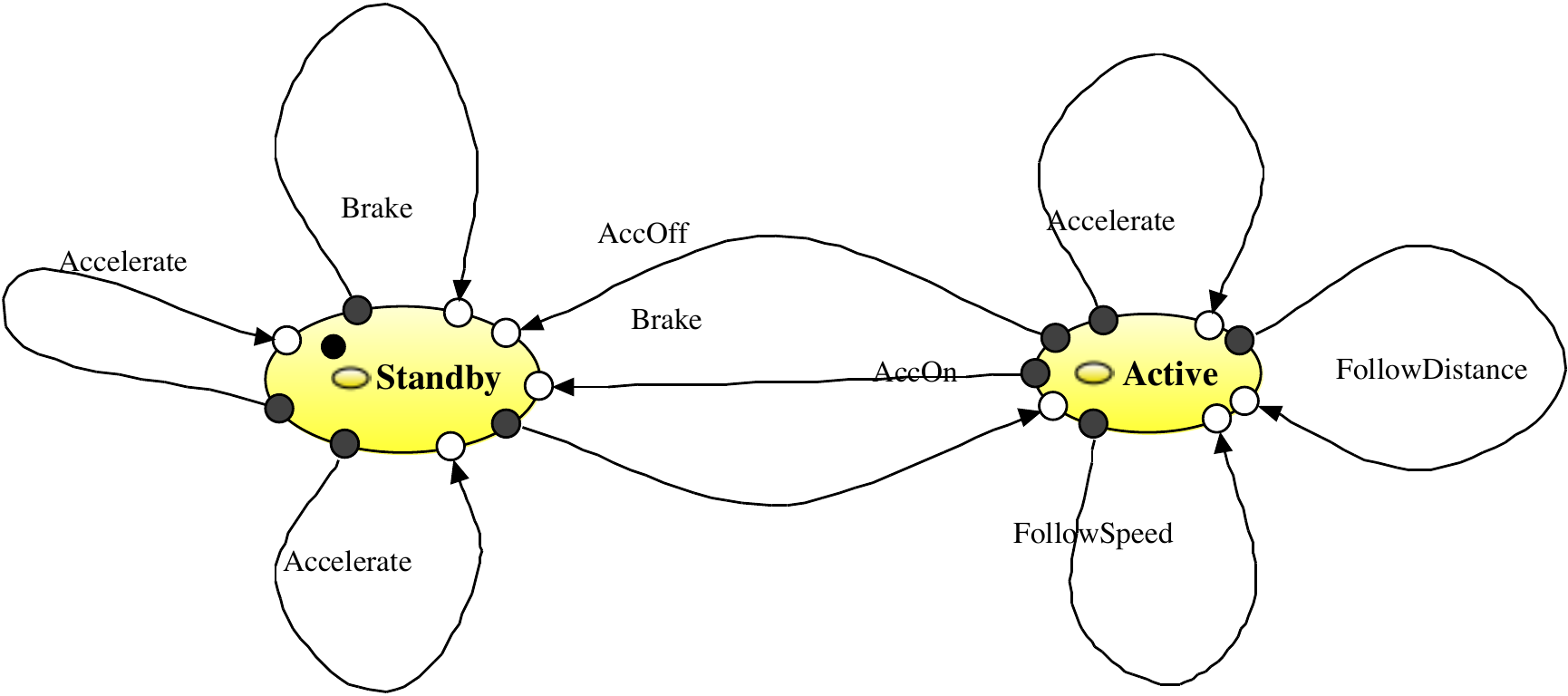}
\end{center}
\caption{An abstract specification}
\label{fig:acc_state-af}
\end{figure}

One high-level behavioral specification of  the  acceleration control component is shown in Figure~\ref{fig:acc_state-af} and also modeled in AutoFocus 3. It consists of two states.  The left state represents the Standby-Mode and the right state is for Active-Mode. The 3 transitions between the states represent the possible state switch of the component. The driver can turn on the ACC (Active-Mode) via a switch, which is represented by the AccOn transition, or turn off the ACC (Standby-Mode) any time by explicitly using the switch, which is the AccOff transition, or implicitly giving a manual brake command, which is the Brake transition. If the component is in the Active-Mode, the ACC system will control the car to follow the given speed and distance rules, if this is for some reason not desirable it does the acceleration according to the driver's command. If it is in the Standby-Mode, it only reacts on the driver's commands.

\subsection{ACC Example Requirements}
In the ISO 15622 standard requirements are explicitly given in the informal text format. 

Here, we present two example requirements which we discuss in more details:

\begin{enumerate}
\item 
To ensure the driver can override the ACC, a requirement about the reaction of the ACC on the brake event is specified: \\
\begin{it}
Braking by the driver shall deactivate the ACC function at least if the driver initiated brake force demand is higher than the ACC initiated brake force.
\end{it}
\item
Requirements can also imply constraints on relations between components: \\
\begin{it}
When the ACC is active, the vehicle speed shall be controlled automatically either to maintain a time gap to a forward vehicle, or to maintain the set speed, whichever speed is lower. The change between these two control modes is made automatically by the ACC system.
\end{it}
\end{enumerate}
Other requirements posses a similar level of complexity.

\begin{figure} 
\begin{center}
\includegraphics[width=0.55\textwidth]{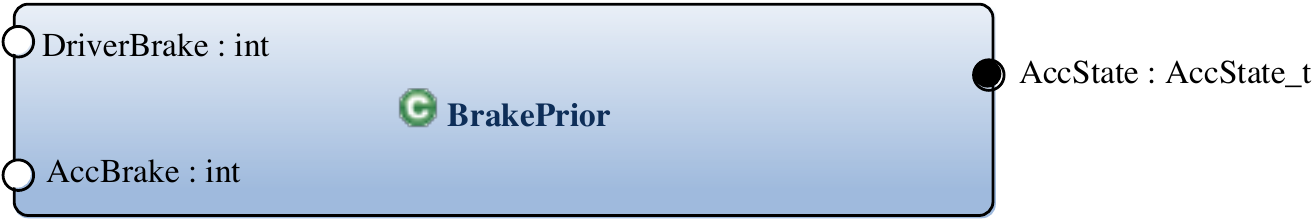} 
\end{center}
\begin{center}
	\includegraphics[width=0.55\textwidth]{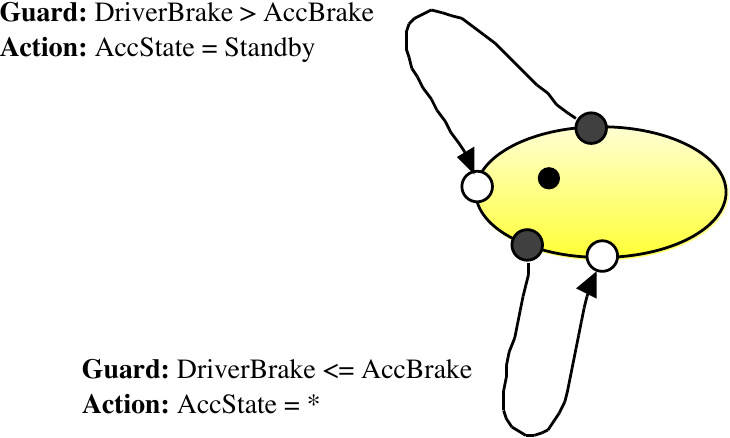}
\end{center}
\caption{An ACC requirement in \focus}
\label{fig:acc-af}
\end{figure}

In our approach, the requirement must first be formalized as a model. 
Figure~\ref{fig:acc-af} shows our first ACC requirement as an AutoFocus 3 component.
In the upper part, the abstract component is shown, the lower part defines a simple automaton. The abstract model has less input streams: {\it DriverBrake} indicating whether the brake padel is down and to what degree, {\it AccBrake} indicating the braking value currently calculated by some other ACC Components and AccState indicating whether the ACC shall currently actively control the car's speed.
The first transition on the top left represents that the ACC system must be automatically set to Standby-Mode, whenever the driver gives brake command and the value is larger than the value computed by the ACC system. If the driver's brake value is smaller than the value computed by the ACC system or the driver gives no brake command, the system state is not concerned.

The second example requirement can be modeled with an AutoFocus component with two input ports $in_{distance}$, $in_{speed}$ and one output port. The two input ports represent suitable acceleration values for maintaining a time gap ($in_{distance}$) and for maintaining speed set by the driver ($in_{speed}$). The output port represents the selected acceleration value according to the inputs. The internal behavior of the component can be described with just one formula: $out = min(in_{distance}, in_{speed})$.

\subsection{Concretization of a Test-case}
An example test-case about the requirement from Figure~\ref{fig:acc-af} comprises streams for 
\begin{center}
DriverBrake $[21,51,78,100,91]$, AccBrake $[79,100,100,91,51]$ 
\end{center}
and the corresponding stream for AccState 
\begin{center}
$[Active,Active,Active,Active,Standby]$.
\end{center}
Output of the stream processing function is delayed by one time step.
\paragraph{Relating abstract and concrete models}
Our formalizations of the $f$ and $f^{-1}_p$ (for different parameters) functions and the {\sf RI} and {\sf RO} relations realize the following constraints:
\begin{itemize}
\item Per time tick, the values of  {\it AccBrake} are either equal to {\it ReqDistAcc} or {\it ReqSpeedAcc} (if we take the second requirement at this stage into account, we already know that it is advisable to chose the least of the two values per time tick),
\item {\it AccMode} does not have a correspondence in the abstract domain an thus does correspond to a parameter values,
\item {\it BrakeCmd} in the concrete domain does correspond to the {\it AccState} stream in the abstract domain: the ACC must be switched to stand-by mode if a certain brake force is applied,
\item {\it GasCmd} corresponds to a parameter value.
\end{itemize}
The $g$ function is less restrictive than  $f^{-1}_p$, e.g., most parameterized values may correspond to any possible parameter.

\paragraph{Comparing abstract and concrete test-results}
In our case study we created an AutoFocus component that abstracts the results of concrete case study output streams element wise. These are compared to the results of the abstract case study and a corresponding boolean value is emitted. Thus, we are computing $g^{-1}$ and in some cases $g$.

\subsection{Implementation and Evaluation}
\label{sec:impl}
\begin{figure} 
\begin{center}
	\includegraphics[width=0.99\textwidth]{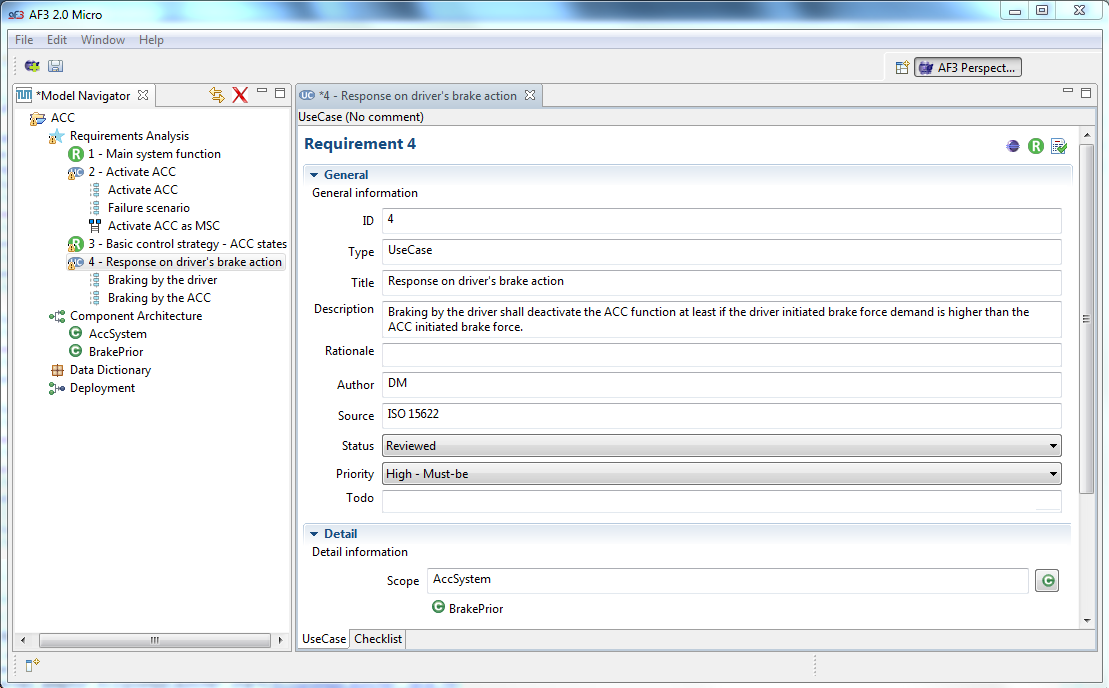}
\end{center}
\caption{Textual requirement in AutoFocus 3}
\label{fig:mira}
\end{figure}

\begin{figure} 
\begin{center}
	\includegraphics[width=0.99\textwidth]{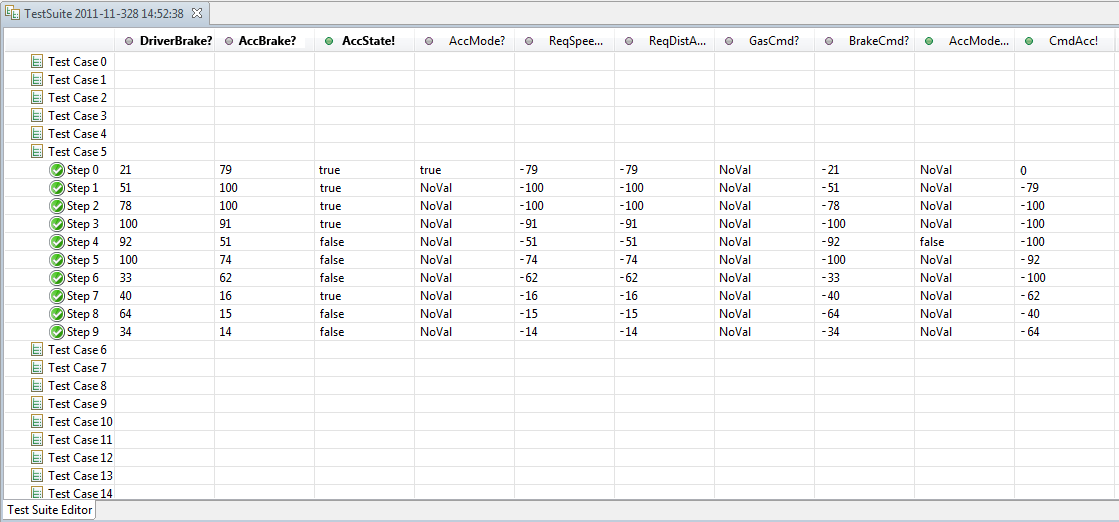}
\end{center}
\caption{Handling a test-case in AutoFocus 3}
\label{fig:aftestcasehandling}
\end{figure}

We have implemented a test-case generation and an abstraction framework for \focus and evaluated it using the ACC case-study described in Section~\ref{sec:acc}.
AutoFocus 3 allows assigning models to textual requirements. A screen-shot of handling the first example requirement in our tool is shown in  Figure~\ref{fig:mira}. 
Figure~\ref{fig:aftestcasehandling} shows a screen-shot of our plug-in to specify and handle abstract  test-cases and their concretization. Once one has specified components that realize $f^{-1}_p$ and $g^{-1}$ or $g$ test cases can be handled in a user-friendly way: on the left side, one can select a test-case, which comprises values for different time steps, thus each column specifies an input stream.

Our implementation works together with an existing test-case generation tool for AutoFocus 3 described in \cite{pfaller}. This test-case generator uses Eclipse based constraint logic programming to derive test-cases.  The test-case generator can be used to generate test-cases for the  abstract model. Furthermore, the implementation can also be used to derive a family of different concrete test cases for a single abstract test-case.

Regarding our framework described in Section~\ref{sec:framework}, for a requirement, we are stating  the {\sf RI}, {\sf RO} relations from  Section~\ref{sec:correctnessRIRO} or the Galois connection from Section~\ref{sec:galois} formally using a mathematical definition first. In case of using the Galois connection approach, we are implementing the  $f^{-1}_p$ (parameterized)    and $g$ functions for test-case abstraction by writing AutoFocus components that realize these functions or the required parts. It is important to notice that we can not simple replace $f^{-1}_p$ by $g$ or vice versa since $f^{-1}_p \subseteq g$ holds in general. The implemented components realizing  $f^{-1}_p$ and $g$ in the regarded cases are relatively simple.  In particular it is possible to derive $f$ from $f^{-1}_p$. It is thus possible to verify the Galois connection property, thereby, guaranteeing that the concrete model is a refinement of the abstract model.

However, we do not have a formal framework for carrying out correctness proofs yet. Such a framework could be established using the semantics of AutoFocus and its formalization in a higher-order theorem prover and remains subject to future work.

\section{Conclusion}
\label{sec:conc}
In this paper we presented a framework for relating components at different abstraction levels with each other. We regarded transforming existing test-cases for abstract models to more concrete models and applying them to these concrete models in the context of our model-based development environment and tool AutoFocus 3. We described an evaluation of the approach using an ACC based case study. Our experiences indicate that reusing test-cases for different abstraction levels in a seemless model-based development process is feasible for software development in embedded systems.

Different directions are interesting for future work.
As next steps, we are interested to take more benefit from our work on relating abstract and concrete models using Galois connections. The investigation of compositionality aspects of tests for different components in a system is a goal.
We are also interested in transforming other properties like, e.g., invariants from an abstract model to a concrete one (cf. \cite{BHH11} for a similar approach that we investigated).

As a more challenging long term goal, we are interested in automatically generating the AutoFocus components that relate abstract and concrete models with each other. Although this is an undecidable problem, some heuristics for relatively simple models would be a great benefit. An additional goal is a further improvement of the quality of test-case  generation in AutoFocus.

\bibliographystyle{eptcs}
%%%\bibliography{biblio}

\end{document}